# Spatiotemporal Superfocusing

Qianru Yang[1,†], Haotian Wu[1,†], Hao Hu[2], F. J. García-Vidal[3], Guangwei Hu[1*], Yu Luo[2*]

[1] School of Electrical and Electronic Engineering, Nanyang Technological University, 50 Nanyang Avenue, Singapore 639798, Singapore.

[2] National Key Laboratory of Microwave Photonics, Nanjing University of Aeronautics and Astronautics, Nanjing 211106, China.

[3] Departamento de Física Teórica de la Materia Condensada and Condensed Matter Physics Center (IFIMAC), Universidad Autónoma de Madrid, Madrid E-28049, Spain.

[†] Q. Yang and H. Wu contributed equally

*Corresponding authors. *Email*: guangwei.hu@ntu.edu.sg (G. Hu); yu.luo@nuaa.edu.cn (Y. Luo)

Superfocusing confines light within subwavelength structures, breaking the diffraction limit. Structures with spatial singularities, such as metallic cones, are crucial to enable nanoscale focusing, leading to significant advancements in nanophotonics, sensing, and imaging. Here, we exploit the spatiotemporal analogue of the wedge structure, i.e. a dielectric medium sandwiched between two subluminal interfaces with distinct velocities, to focus propagating waves beyond the diffraction limit, achieving spatiotemporal superfocusing. Within this structure, an incident pulse undergoes continuous spatial and temporal compression due to Doppler effects, which accumulates and results in an extreme focusing as it approaches the spatiotemporal vertex. Remarkably, unlike the field localization in conventional superfocusing, the compressed light in spatiotemporal wedges experiences significant amplification and then couple to the far field in free space. Our findings represent an indispensable paradigm for extreme concentration and amplification of propagating waves in space-time dimensions.

# Introduction

The resolution of an optical imaging system is constrained by the Abbe diffraction limit, which also prevents the focusing of light within a subwavelength region. As illustrated in Fig. 1(a), the adiabatic narrowing of the waveguide width leads to an increasingly delocalized electric field distribution. To overcome the diffraction limit, surface plasmon polaritons (SPPs) have been extensively investigated and applied in optical imaging [1], sensing [2], and spectroscopy [3-5]. As depicted in Fig. 1(b), the wavelength of the excited SPP can be progressively compressed during propagation towards the tip, ultimately allowing the deep subwavelength focusing of light beyond the diffraction limit, a phenomenon known as *superfocusing*. This superfocusing shows extreme field enhancement and hence significantly enhanced the strong light-matter interaction for extreme ultraviolet (EUV) generation [6] and hot-electron conversion with high efficiency [7] and for fundamental study of the photoelectron behavior in the strong-field regime [8].

The realization of superfocusing down to the nanoscale relies on the structures with spatial singularities, such as metallic wedges [9], cones [10], and dimers [11]. Theoretically, SPPs can be compressed into a singular point without reflection enabling both the propagation constant and field intensity to diverge even when the absorption of metal is considered [9], analogous to an optical black hole, where all light within the event horizon is drawn towards the singular point [12]. Spatial resolution down to ~ 5 nm has been experimentally demonstrated through the metallic wedge-assisted superfocusing [13]. Notably, such experiment requires extremely high manufacturing precision, as any deviation from the singular point can lead to significant energy losses (exceeding 90%) due to limited SPP coupling efficiency, dissipative losses, and reflection [14]. As time-reversal symmetry is preserved in non-magnetic static structures, the forward and backward light propagation exhibits identical physical characteristics, thereby rendering reflection inevitable.

In recent years, active artificial materials with both spatial and temporal modulations have garnered significant attention for their unprecedented control of electromagnetic waves [15-21]. Particularly, harnessing the time varying and modulation as a new degree of freedom, various applications have been employed, such as amplification [22-24], frequency generation [25-27], frequency compensation for phase matching [28-31], exploring relativistic phenomena [32], and breaking bandwidth limits in passive systems for impedance matching [33]. Recently, researchers have identified a novel gain mechanism [34-37]. Studies have demonstrated that a grating modulated at the speed of light exhibits a series of discrete gain points, analogous to black holes [37, 38]. For an infinitely long grating, both the frequency and the electric field intensity diverge. Moreover, the breaking of time-reversal symmetry results in a unidirectional and nonreciprocal amplification effect, ensuring robustness against reflections. Similar singularities of fields can be realized through scattering by the interluminal boundary [39].

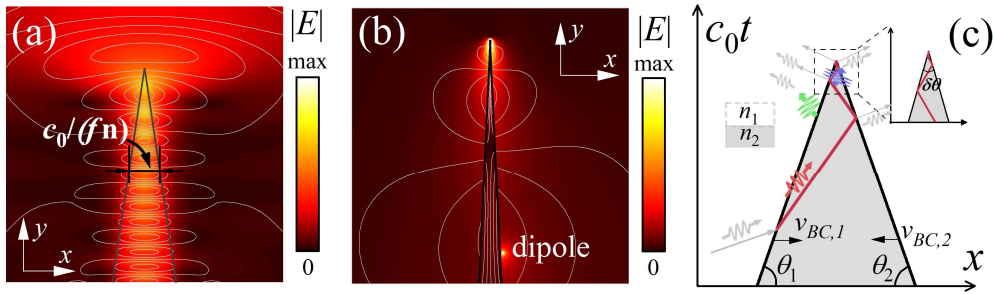

**Fig. 1**. (a) Directing the fundamental guided mode to a sharp dielectric wedge with refractive index $n$ at the frequency of $f$. (b) Superfocusing on a metallic wedge. The colors in (a) and (b) indicate the magnitude of the electric field. The gray solid lines represent the contour lines of the real part of $E_z$ in (a) and the real part of $H_z$ in (b). (c) Schematic of superfocusing in a spatiotemporal wedge.

Inspired by the singularities in interluminal moving gratings and boundaries, we propose the concept of *spatiotemporal superfocusing*, which involves the amplification of optical frequency and energy in spatiotemporally modulated systems. Our wedge in the space-time domain can replace the interluminal modulation, demonstrating superfocusing at lower modulation speeds and enabling controllable growth of frequency and energy. As shown in Fig. 1(c), the wedge is composed of two intersecting subluminal moving boundaries and the light beam experiences endless cascaded reflections in an ideal untruncated spatiotemporal wedge. We identified the critical angles necessary for achieving spatiotemporal superfocusing and an analysis of the impact of tip bluntness on wave confinement and amplification dynamics. Notably, the proposed spatiotemporal superfocusing mechanism compresses the electromagnetic field in both space and time while preserving its propagation characteristics. This intriguing aspect distinguishes itself from plasmonic superfocusing, which relies on the inherently decaying nature of SPPs. Our model extends the concept of superfocusing to spatiotemporal domain and expands the ability for electromagnetic wave control.

## Results and Discussions
### Superfocusing on the spatiotemporal wedge

In this study, we consider a spatiotemporal wedge surrounded by two intersecting moving boundaries. Due to practical limitations in spatiotemporal resolution, it is assumed that the tip of the wedge is truncated by a temporal boundary, expressed as:

$$\varepsilon(x,t) = \varepsilon_1 + (\varepsilon_2 - \varepsilon_1)\left[u(x + \delta/2 - v_{BC1}t) - u(x - \delta/2 - v_{BC2}t)\right]u(-t), \qquad (1)$$

where $u(x - x_0)$ denotes a step function that jumps from 0 to 1 at $x_0$, $\delta$ is the tip width of the truncated wedge. The velocities of the first and second moving boundaries are $v_{BC1}$ and $v_{BC2}$, respectively, as indicated in Fig. 1(c). The phase velocity of waves in medium 1(or 2) is denoted as $v_{1(2)} = c_0(\varepsilon_{1(2)}/\varepsilon_0)^{-1/2}$. In this study, we only consider the subluminal modulations, i.e. $|v_{BC1(BC2)}| < v_1, v_2$; otherwise, the beam will be surpassed by boundaries and cannot be confined within the wedge through multiple reflections. Note that there is no physical movement of the media, rather the variation of the permittivity between $\varepsilon_1$ and $\varepsilon_2$ to produce the moving interfaces.

To begin, we briefly investigate the scattering of waves by a subluminal moving boundary between two media with velocity $v_{BC}$. Here, we define a moving frame using Lorentz transformation: $x' = \gamma\left(x - \frac{v_{BC}}{c_0}c_0 t\right)$ and $c_0 t' = \gamma\left(c_0 t - \frac{v_{BC}}{c_0}x\right)$, where $\gamma = 1/\sqrt{1 - v_{BC}^2/c_0^2}$ [40]. In spacetime coordinate $x$-$c_0 t$, it rotates the $x$-axis with angles $\theta = \arctan(v_{BC}/c_0)$, defining the rapidity of the moving frame [41]. Due to the breaking of translational symmetry in space and time, $k$ and $\omega$ are not conserved. This results in the Doppler shifts for reflected and transmitted waves, expressed as $\omega_r = \gamma_r \omega_i$ and $\omega_\tau = \gamma_\tau \omega_i$, where $\omega_i$ denotes the incident angular frequency. The two terms $\gamma_r(v_{BC}, v_1) = \frac{1 - v_{BC} v_1^{-1}}{1 + v_{BC} v_1^{-1}}$ and $\gamma_\tau(v_{BC}, v_1, v_2) = \frac{1 - v_{BC} v_1^{-1}}{1 - v_{BC} v_2^{-1}}$ are denoted as the reflection and transmission scaling factors, respectively. Meanwhile, the reflection and transmission coefficients are also scaled relative to their static counterparts, given by $r = \gamma_r r_0$ and $\tau = \gamma_\tau \tau_0$, where the transmission and reflection coefficients for the static spatial boundary are expressed as $\tau_0 = \frac{2n_1}{n_1 + n_2}$ and $r_0 = \frac{n_1 - n_2}{n_1 + n_2}$. To show the energy exchange through these time-varying media, we further evaluate the Poynting theorem. Energy transfer into and out of the system due to a moving boundary can be expressed as $P_G = -2v_{BC}\varepsilon_1 \frac{n_2 - n_1}{n_2 + n_1} \frac{(1 + v_{BC} v_2^{-1})(1 - v_{BC} v_1^{-1})}{(1 - v_{BC} v_2^{-1})(1 + v_{BC} v_1^{-1})} P_i$, where $P_i$ is the power flow of incident wave.

The total power flow is amplified via the moving boundary as the permittivity (capacitance) transitions from a higher value to a lower one (e.g., $v_{BC} < 0$, $n_2 > n_1$). A similar situation arises when, with current or charge remaining constant, a decrease in capacitance—such as suddenly increasing the distance between capacitor plates—will results in an increase in the energy of the LC circuit.

Next, we examine the scattering of fields by a spatiotemporal wedge. As shown in Fig. 1(c), from a geometric perspective, the trajectories of pulses and the boundaries of the wedge form a sequence of similar triangles in spacetime coordinates, resulting in indefinitely sustained multiple reflections. As a result, the frequencies and amplitudes may diverge in a cascaded manner. The angular frequency and the amplitude for forward and backward propagating waves in the wedge can be written as $\omega_{F(B)}^{(N)} = \gamma_{F(B)}^{(N)} \omega_i$ and $|E_{F(B)}^{(N)}| = \gamma_{F(B)}^{(N)} |E_{F(B),0}|$, with the scaling factors given by,

$$\gamma_F^{(N)} = \gamma_{\tau,BC1} \left( \gamma_{r,BC2} \gamma_{r,BC1} \right)^N \gamma_{\tau,BC2}, \tag{2a}$$

$$\gamma_B^{(N)} = \begin{cases} \gamma_{r,BC1}, & N = 0 \\ \gamma_{\tau,BC1} \gamma_{r,BC2} \left( \gamma_{r,BC1} \gamma_{r,BC2} \right)^N \gamma_{\tau,BC1}, & N \geq 1 \end{cases}, \tag{2b}$$

where the subscripts $F$ and $B$ indicate the forward and backward propagating waves and the superscript $N$ denotes the order of the roundtrip scattering in the spatiotemporal wedge and $N \in \mathbb{N}_0$. The scaling factors for reflected and transmitted waves of first and second moving boundaries are given by $\gamma_{r,BC1(BC2)} = \gamma_r(v_{BC1(BC2)}, v_2)$, $\gamma_{\tau,BC1(BC2)} = \gamma_\tau(v_{BC1(BC2)}, v_2, v_1)$. The corresponding amplitudes of scattered fields due to static boundaries are given by $|E_{F,0}^{(N)}| = \tau_0^2 (r_0)^{2N} |E_i|$ and $|E_{B,0}^{(N)}| = \begin{cases} r_0 |E_i|, & N = 0 \\ \tau_0^2 r_0^{2N+1} |E_i|, & N \geq 1 \end{cases}$. To compress and amplify light beam, it requires $v_{BC1} > v_{BC2}$, which is the necessary condition for $\gamma_{r,BC1} \gamma_{r,BC2} \geq 1$; hence, only if a spatiotemporal wedge points in the direction of time flow can it achieve spatiotemporal superfocusing. Meanwhile, based on the Poynting's theorem, $\varepsilon_2 > \varepsilon_1$ is required for amplification after a roundtrip.

In the case of symmetric wedges with $v_{BC1} = -v_{BC2}$, the reflection scaling factor is always greater than 1, i.e., $\gamma_r(v_{BC1}, v_2) = \gamma_r(v_{BC2}, v_2) = \gamma_r > 1$, leading to increasing reflection frequencies and amplitudes. The total field is the superposition of all light beams scattered by the spatiotemporal wedge. We consider scattering from a truncated spatiotemporal wedge expressed as Eq. (1). As shown in Fig. 2(a), a Gaussian pulse impinges on a spatiotemporal wedge, which results in two sharp and strong pulses re-radiating along the forward and backward directions. The enlarged field distribution near the tip is depicted in the inset of Fig. 2(a). To clearly show the pulse evolution, we depict a series of snapshots in Fig. 2(b). As the thickness of the wedge uniformly decreases from $t_1$ to $t_3$ (indicated by the gray area in Fig. 2(b)), the pulses are continuously compressed and amplified. At time $t_4$, two pulses radiate into the far-field regime along opposite directions, as shown in the bottom panel of Fig. 2(b).

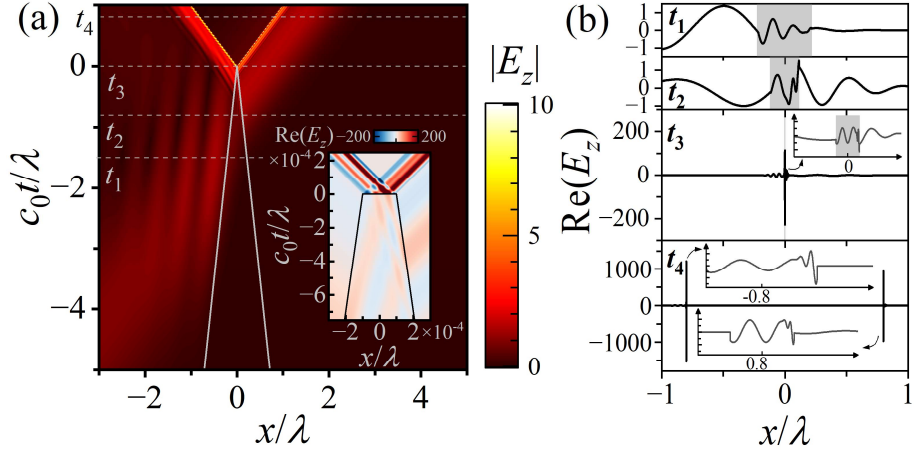

**Fig. 2**. (a) The electric field distribution for a pulse scattered by a spatiotemporal wedge. Inset: The enlarged field distribution near the tip, where the spatiotemporal wedge is truncated by a temporal boundary. (b) Snapshots of field distribution in (a) from $t_1$ to $t_4$. The gray region indicates the spatiotemporal wedge whose width decreases as time evolves. The insets in (b) show the enlarged waveform of the pulses. The trajectories of the moving boundaries are depicted by gray solid lines. Here, the wedge is truncated at $t = 0$ with the tip width of $\delta = 2\times10^{-4}\lambda$. The permittivities of media $\varepsilon_1 = 1$ and $\varepsilon_2 = 3.5^2$. The velocities of the moving boundaries are $v_{BC1} = -v_{BC2} = v_2/2$ and the opening angle of the wedge is $\delta\theta = 16.26°$. The incident pulse has a unit peak amplitude.

The unique superfocusing phenomena between metallic wedges and spatiotemporal wedges can be explained by their distinct wave compression mechanisms. For plasmonic superfocusing, the fields are localized at the nanoscale with conserved frequency. However, the proposed spatiotemporal wedge compresses the wave in both space and time, thereby retaining its propagation characteristics. The evanescent characteristics of the former limit its applicability to only the near field and require additional optical components for coupling. In contrast, the latter offers a novel superfocusing mechanism, preserving propagation characteristics that ensure compatibility with modulated waveguides [42, 43], making it suitable for integrated platforms.

## Critical conditions for spatiotemporal superfocusing

In the case of an untruncated spatiotemporal wedge, if the reflection coefficient for each round trip exceeds one, i.e., $\gamma_r(v_{BC1},v_2)\gamma_r(v_{BC2},v_2) \geq r_0^{-2}$, the field intensity will exponentially increase over time and diverge at the tip. We refer to this phenomenon as *spatiotemporal superfocusing*. At the critical threshold, where the equation holds as equality ($\gamma_{r,BC1}\gamma_{r,BC2} = r_0^{-2}$), the superposed scattered waves with equal amplitudes can also lead to a diverging field intensity. Conversely, below this threshold, the superposition of scattered fields will be amplified but eventually converge to a finite value, which we define as *spatiotemporal focusing*. Compared to luminal regimes, wave compression and energy extremes based on spatiotemporal wedges are controllable. In the following sections, we will illustrate the critical condition for spatiotemporal superfocusing with respect to the geometry of wedges in spacetime coordinates, including the opening angle $\delta\theta = \pi - (\theta_1 + \theta_2)$ and orientation angle $\bar{\theta} = \dfrac{\theta_1 - \theta_2}{2}$, where the expressions of rapidity for moving boundaries are expressed as

$$\theta_1 = \pi u(-v_{BC1}) + \mathrm{atan}\frac{c_0}{v_{BC1}} = \frac{\pi}{2} + \left(\bar{\theta} - \frac{\delta\theta}{2}\right) \quad \text{and} \quad \theta_2 = \pi u(v_{BC2}) - \mathrm{atan}\frac{c_0}{v_{BC2}} = \frac{\pi}{2} - \left(\bar{\theta} + \frac{\delta\theta}{2}\right).$$

First, we consider the critical condition for superfocusing in the symmetric case with $\bar{\theta} = 0$. The critical opening angle for superfocusing can be determined by $\gamma_r = r_0^{-1}$, which can be expressed as:

$$\delta\theta_c = \pi - 2\arctan\frac{n_2^2}{n_1}. \tag{3}$$

As $\delta\theta$ increases, the scaling factor $\gamma_r$ monotonically increases, leading to spatiotemporal superfocusing when $\delta\theta \geq \delta\theta_c$.

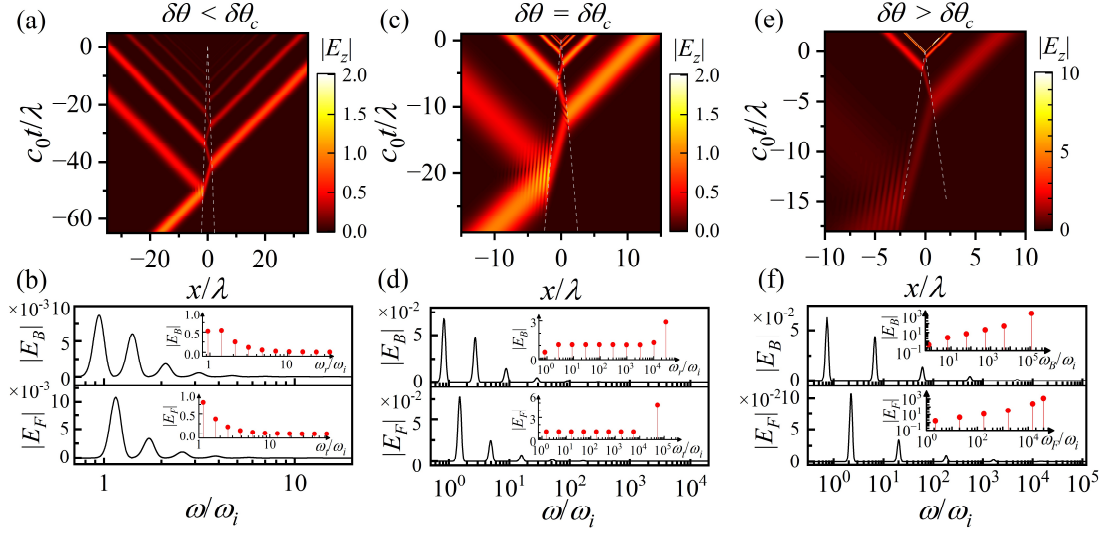

**Fig. 3**. Critical phenomena for spatiotemporal superfocusing demonstrated by multi-scattering from wedges. The magnitude of electric field $|E_z|$ in spacetime coordinates with the opening angle (a) ($\delta\theta = 3.27°) < \delta\theta_c$; (c) $\delta\theta = (\delta\theta_c = 9.33°)$; (e) ($\delta\theta = 16.26°) > \delta\theta_c$. The spectra of fields at positions of $x = 10\lambda$ and $x = -10\lambda$ (denoted as $E_F$ and $E_B$) are shown in (b), (d), and (f), respectively. The insets show the integral of each Gaussian pulse versus its central frequency. Here, $\delta = 2\times10^{-4}\lambda$.

To illustrate the critical effect of superfocusing, we consider a pulse incident from a position far from the tip. When $\delta\theta < \delta\theta_c$, the amplitude of the pulse decreases with each reflection and dissipates as it approaches the tip, as illustrated in Fig. 3(a). The spectra of the scattered pulses at $x = 10\lambda$ and $x = -10\lambda$ are shown in Fig. 3(b). After each scattering event, the pulse undergoes compression, leading to a broadening of pulse width by a factor of $1/\gamma_{\tau(r)}$. The inset presents the integrated area of each pulse relative to its central frequency, where the pulse area is proportional to the scattering coefficient for monochromatic wave excitation. When $\delta\theta = \delta\theta_c$, the reflection coefficient within the spatiotemporal wedge reaches unity. Aside from the first backward-propagating pulse and the wave scattered by the truncated wedge, all scattered fields exhibit the same amplitude, as seen in Fig. 3(c). Due to the broadening of the pulse width, the peak amplitude of each pulse decreases, as depicted in Fig. 3(d), while the pulse area remains unchanged, as shown in the inset. By further increasing the opening angle, the intensity of the scattered waves is significantly amplified by repeated reflections, as demonstrated in Fig. 3(e). The corresponding spectrum is shown in Fig. 3(f), where both the frequency and amplitude grow exponentially, as indicated in the inset of Fig. 3(f).

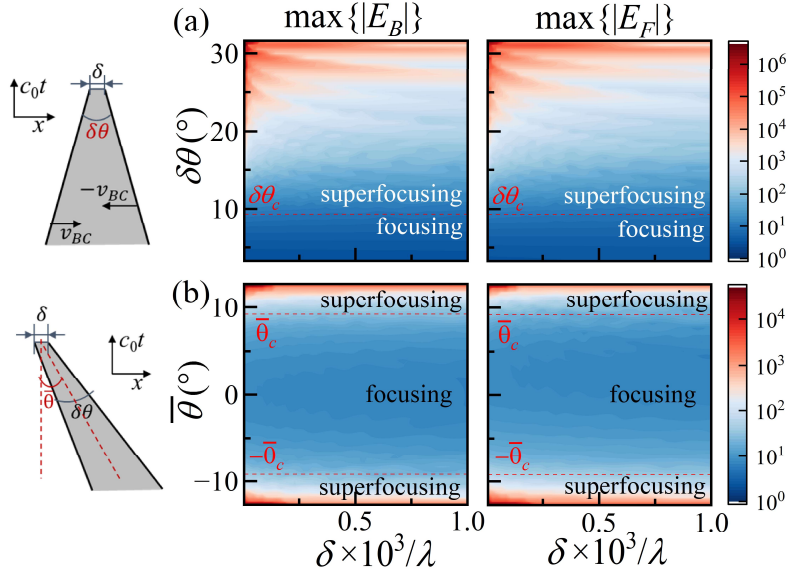

**Fig. 4**. Dependencies of the maximum amplitudes of forward and backward propagating pulses on the opening angle (a) and the orientation angle (b). The left panel depicts the geometry of the spatiotemporal wedges. Here, $\bar{\theta} = 0°$ in (a) and $\delta\theta = 6°$ in (b).

Next, we evaluate the superfocusing condition for the orientation angle. In contrast to a spatial wedge, a spatiotemporal wedge does not possess rotational symmetry, because time flows in a single direction toward the future. The rotation of a wedge in spatial coordinates is characterized by the SO(2) group, while for spatiotemporal wedges, it is represented by the Lorentz group $SO^+(1,1)$ [41]. By fixing the opening angle and increasing the magnitude of the orientation angle $\bar{\theta}$, the scaling of the reflection coefficient changes monotonically, resulting in a transition from focusing to superfocusing. The critical orientation angle is determined by the following analytical expression:

$$\bar{\theta}_c = \mathrm{acot} \sqrt{\frac{\left(n_2 \cot\dfrac{\delta\theta}{2}+\gamma_+\right)^2 - \gamma_-^2}{\left(\cot\dfrac{\delta\theta}{2} - n_2\gamma_+\right)^2 - \left(n_2\gamma_-\right)^2}}, \qquad (4)$$

where $\gamma_+ = \dfrac{1}{2}\left(\dfrac{n_1}{n_2} + \dfrac{n_2}{n_1}\right)$ and $\gamma_- = \dfrac{1}{2}\left(\dfrac{n_1}{n_2} - \dfrac{n_2}{n_1}\right)$.

To demonstrate the superfocusing effects and their critical conditions, we examine the dependence of the maximum scattered amplitudes of forward and backward-propagating waves on the tip width ($\delta$) and the opening/orientation angles. For the symmetric case ($\bar{\theta} = 0$), when $\delta\theta \geq \delta\theta_c$, the maximum amplitude of the superposed scattered fields increases as the tip narrows, leading to extreme field amplification and exhibiting the characteristics of a superfocusing effect. Otherwise, the scattered fields initially undergo amplification but eventually stabilize at a maximum value, as shown in Fig. 4(a). Similar critical phenomena are observed when increasing $|\bar{\theta}|$, as presented in Fig. 4(b). As the orientation angle increases towards the interluminal regime, the reflection scaling factor tends to infinity, as described by Eqs. 2(a) and 2(b). For $\bar{\theta} \neq 0$, the distribution of maximum amplitudes becomes asymmetric with respect to $\bar{\theta} = 0$, indicating the nonreciprocal characteristics of scattering from spatiotemporal wedges.

## Influence of blunted tip

In practical experiments, the non-instantaneous variation in optical parameters is a non-negligible factor, leading to a blunted tip, which truncates cascaded reflections and reduces the reflection coefficient. We calculate the corresponding scattered fields using FDTD numerical simulations. The blunted spatiotemporal wedge can be expressed as:

$$\varepsilon(x,t) = \varepsilon_1 + \frac{1}{2}(\varepsilon_2 - \varepsilon_1)\left[\tanh\left(k_{BC}(x - v_{BC}t)\right) - \tanh\left(k_{BC}(x + v_{BC}t)\right)\right]u(-t), \qquad (5)$$

where $k_{BC}$ represents the sharpness of the moving boundary. The permittivity profile of the wedge, as described by Eq. (5), is depicted in Fig. 5(a), and the scattered fields are shown in Figs. 5(b) and 5(c). The scattered fields illustrate that a decrease in sharpness, i.e., $k_{BC}$, leads to a significant reduction in the intensity of the backward-scattering waves, while the forward-propagating waves remain unchanged. These phenomena are more clearly observed in the corresponding spectra, as illustrated in Figs. 5(d) and 5(e). Specifically, for the backward-scattered pulses, the amplitude of the high harmonic components rapidly decreases, accompanied by a noticeable redshift compared to the ideal wedge as $k_{BC}$ approaches infinity. In contrast, the low-frequency components remain stable. Due to the significant role of high-frequency component intensity, the blunted wedge markedly affects the performance of spatiotemporal superfocusing, particularly in terms of backscattering.

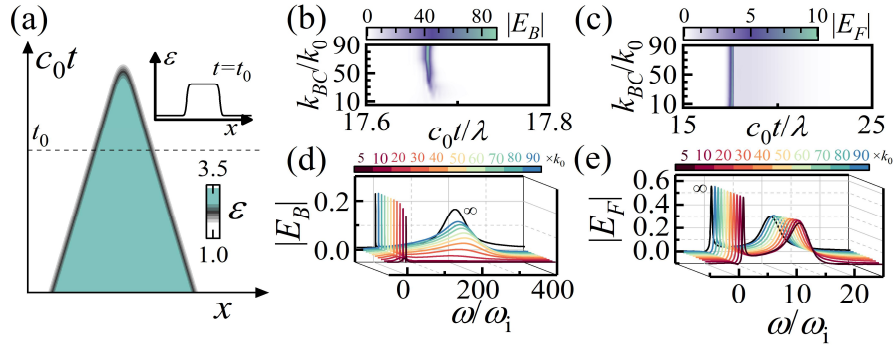

**Fig. 5**. The effect of smoothed boundaries on spatiotemporal superfocusing. (a) Profile of permittivity for a rounded spatiotemporal wedge and the distribution of the index at time $t_0$ (inset). (b) & (c) The backward and forward propagating pulses scattered by the rounded wedge, respectively. (d) & (e) The corresponding spectra in (b) and (c). Here, $\delta\theta = 28.84°$; the fields of backward (forward) propagating waves are recorded at $x = -4.5\lambda$ ($4.5\lambda$).

The experimental realization of spatiotemporal wedges remains challenging, particularly in achieving rapid and strong modulation, as well as independent control over temporal and spatial dimensions. Recently, FPGA-controlled two-dimensional diode arrays was employed to generate microwave spatiotemporal coding metasurfaces [44]. Subsequently, temporal boundaries have been experimentally demonstrated using transmission line metamaterial through voltage-controlled reflective switches [42], further showcasing the potential for extending towards the realization of moving boundaries and spatiotemporal wedges. On the other hand, at telecom wavelengths, electrically driven waveguides equipped with distributed PIN diodes was employed to realize nonreciprocity [43]. Although the wave velocity within waveguides is below the speed of light, realizing subluminal spatiotemporal boundaries and wedges remains challenging. Nevertheless, the existing experimental platforms present promising pathways toward achieving spatiotemporal wedges. The spatiotemporal superfocusing mechanism proposed in this paper retains propagation characteristics, demonstrating unique advantages and irreplaceable potential for integrated applications and subsequent signal processing.

## Conclusion

We have constructed a wedge configuration in spacetime coordinates to achieve spatiotemporal superfocusing for the first time, allowing light beams to be compressed, amplified, and re-radiated into free space. Unlike near-field superfocusing based on SPs, the pulses scattered by the proposed spatiotemporal wedge remain in propagating modes, with their frequencies and amplitudes scaled due to the relativistic effects of moving boundaries in a cascaded manner. By modifying the geometry of the wedge, specifically the opening and orientation angles, the strength of superimposed fields can transition from convergent to divergent values, corresponding to spatiotemporal focusing and superfocusing, respectively. In practice, for wedges truncated with a flat top, the strength of scattered fields grows exponentially as the tip width becomes sharper, without diverging. Additionally, for blunted tips, we observe a counterintuitive phenomenon where reflections are significantly diminished, while transmission retains its strength even in the presence of pronounced bluntness. Our study provides a novel approach for achieving superfocusing through spatiotemporal wave manipulation, demonstrating distinctive phenomena compared to spatial counterparts and offering additional advantages of being broadband and lossless.

# Acknowledgement

This work is partially sponsored by the Distinguished Professor Fund of Jiangsu Province (Grant No. 1004-YQR24010), Fundamental Research Funds for the Central Universities, NUAA (No. NE2024007), the Singapore National Research Foundation Competitive Research Program (NRF-CRP22-2019-0006 and NRF-CRP23-2019-0007). H.H. acknowledges financial support by National Natural Science Foundation of China (Grant No. 12404363), Fundamental Research Funds for the Central Universities (Grant No. NS2024022), and Distinguished Professor Fund of Jiangsu Province (Grant No. 1004-YQR23064). G. H. acknowledges the Nanyang Assistant Professorship Start-up Grant, Ministry of Education (Singapore) under AcRF TIER1 (RG61/23), and National Research Foundation of Singapore under award no. CRP22-2019-0064.